\documentclass[preprint,amsmath,amssymb]{revtex4}
\usepackage{psfig}

\begin{document}
\title{A procedure for calculating the many-particle Bohm quantum potential.}
\author{L.Delle Site}
 \email{dellsite@mpip-mainz.mpg.de}
\affiliation{Max-Planck-Institute for Polymer Research\\
Ackermannweg 10, D 55021 Mainz Germany.}
\begin{abstract}
In a recent work, M.Kohout
(M.Kohout, {\it Int.J.Quant.Chem. \bf{87}, 12 2002}) 
raised the important question of how to make a correct use
of Bohm's approach for defining a  quantum potential. 
In this work, by taking into account Kohout's results,
we propose a general self-consistent iterative procedure for
solving this problem.
\\
\\ {\bf Key-words}:
\\Bohm's quantum potential, many-electron systems,
partial differential equations, self-consistent procedure.
\end{abstract}
\maketitle

\section{Introduction}
Bohm's formulation of quantum mechanics in terms of single-particle
trajectories (\cite{bohm1,bohm2,bohm3,durr,bohm4,holland}) 
has been, and still is, a continuous matter of dispute, often 
in rather philosophical terms above all 
(\cite{bohm4,holland,bohm4a,bohm5,durr2,durr3,durr4,belousek1,belousek2,matzkin})
concerning the fundamental meaning of quantum
mechanics. At a concrete level such a theory 
 has been used as a theoretical tool for understanding and
 interpreting  several processes in different fields, from molecular physics 
to plasma physics,
from scattering theory to simulation of quantum 
wires \cite{luigieuro,luigiphysa,ham,creon,longo,nerukh,shifren}, to name 
a few. Recently M.Kohout \cite{kohout} raised the important question  
of how to treat the more rigorous and realistic  
$3N$-dimensional formulation  of the quantum potential generated by $N$ electrons.
Rigorously
speaking, in an $N$-particle system Bohm's potential is a
$3N$-dimensional function and the one-particle Bohm's potential,
usually considered in literature, is the simplest $3$-dimensional
reduced form which {\bf systematically} does not take
into account the effects due to the 
presence of other particles. He proposed a formal interpretation of the
wavefunction as a product of a one-particle marginal function 
and a conditional many-particle function where somehow the
effects due to the other particles are taken into account, then 
a formal expression of the quantum potential is obtained.
However, as the author underlines,  {\bf  for the conditional 
many-particle function  an explicit 
expression is required and this is a 
rather difficult problem }. 
In this work, by taking into account  Kohout results we build a 
reasonable initial guess for the many-particle 
potential and, following this choice, we develop  an iterative
self-consistent procedure for obtaining a general (numerical) 
expression of the potential. We restrict our
analysis to a spinless system, 
in any case there exists the possibility to extend the procedure to 
wavefunctions which explicitly consider the spin variables.
To conclude, {\bf we must underline} that the intention of this work 
is simply to show that  the widely used Bohm potential
(used in the approximation of one particle) can actually be reasonably 
treated in its
true form of many-particle and we propose a method to do so. 
{\bf It is not our intention} to prove that such 
a procedure is preferable to other quantum approaches, such as
Hartree-Fock or Density Functional Theory, for determining general
many-body effects in electronic systems.

\section{A brief account of Bohm's theory}
Bohm's formulation of quantum mechanics in terms of single-particle
trajectory is based on the assumption that the wavefunction determines 
the dynamics of more fundamental variables (hidden variables).
The essence of the theory can be summarized by quoting Bohm's original 
work ''{\it The first step in developing this interpretation in a more
  explicit way is to associate with each electron a particle having
  precisely definable and 
continuously varying values of position and momentum}'' \cite{bohm1}.
Within this assumption,
the system is described by its wavefunction $\psi({\bf x}_{1}...,{\bf
  x}_{N},t)$ and the positions of its particles ${\bf x}_{i}$; these two 
fundamental quantities are governed respectively by Schr\"{o}dinger
equation 
\begin{equation}
i\hbar\frac{\partial\psi}{\partial t}= H\psi
\end{equation}
where $H$, as usual, is the Hamiltonian of the system,
and by the dynamical equation:
\begin{equation}
\frac{d}{dt}{\bf x}_{i}={\bf v}_{i}
\end{equation}
where $ {\bf v}_{i}=\frac{\hbar}{m_{i}}Im\frac{\psi^{*}\nabla_{{\bf
      x}_{i}}\psi}{\psi^{*}\psi}$. This procedure leads to a
non-Newtonian
equation of motion which becomes Newtonian in the classical limit of
$\hbar \rightarrow 0$; the connection of such an approach 
with quantum mechanics is provided
by the fact that the quantum formalism automatically emerges from Bohm 
mechanics. In simple terms this approach
''{\it ...implies however the particle moves under
  the action of a force which is not entirely derivable from the
  classical potential, $V({\bf x})$, but which also obtains
  contributions from the quantum mechanical potential...}'' \cite{bohm1}.
Given the $N$-particle electronic wavefunction whose general form can be written as 
$\psi({\bf
  R},t)=\chi({\bf R},t)e^{i\hbar S({\bf R},t)}$; ${\bf R} \in \Re^{3N}$, substituting it into the
time-dependent Schr\"{o}dinger equation and then separating real and
imaginary part, once the velocity is defined as 
${\bf v}=\frac{i\hbar}{2m}
\frac{\psi\nabla\psi^{*}-\psi^{*}\nabla\psi}{R^{2}}=
\frac{1}{m}\nabla S$, and $\rho=|\psi|^{2}=\chi^{2}$ one obtains the
equations:
\begin{equation}
\frac{\partial}{\partial t}\rho+\nabla\cdot(\rho{\bf v})=0
\label{cont}
\end{equation}
and 
\begin{equation}
\frac{\partial S}{\partial t}+\frac{[\nabla S({\bf R},t)]^{2}}{2m}=-\left[V({\bf R},t)+Q({\bf
    R},t)\right].
\label{genmot}
\end{equation}
where $m$ is the mass of the particle, $V({\bf R},t)$ is the 
potential characterizing the system (e.g. electrostatic for
interacting fermions), 
$Q({\bf R},t)=-\frac{\hbar^{2}}{2m}\frac{\nabla^{2}\sqrt{\rho}}{\sqrt{\rho}}$
is the Bohm potential and $\nabla=\sum_{i=1}^{N}\nabla_{i}$ the
sum of the gradients of the $N$-particles.
We restrict our analysis to the stationary case, thus 
we can write
 in terms of the wavefunction phase factor:
\begin{equation}
\frac{[\nabla S({\bf R})]^{2}}{2m}=-\left[V({\bf R})+Q({\bf
    R})\right].
\label{genmot2}
\end{equation}   
\section{Kohout's formulation of the multi-particle potential problem}
Bohm's potential in three dimensions is used in many applications,
within this framework $\rho({\bf r})$, a one-particle electron density,
is defined as:
\begin{equation}
\rho({\bf r})=N\int_{\Omega^{(N-1)}}|\psi({\bf r,
  r_{2},r_{3}......r_{N}})|^{2}
d{\bf{r_{2}}}d{\bf{r_{3}}}....d{\bf{r_{N}}}
\label{dr}
\end{equation}
($\Omega$ domain of definition of the system in real space) and
the single-particle wavefunction writes:
\begin{equation}
\psi({\bf r})=\frac{\sqrt\rho({\bf r})}{N}e^{i\hbar s(\bf {r})};
{\bf r} \in \Re^{3}.
\label{wf}
\end{equation}
The wavefunction's form of Eq.\ref{wf} corresponds to the procedure 
of separating Bohm's dynamical equations 
into independent and indistinguishable single-particle
equations  
and describe a set of identical particles moving in an
average potential where specific mutual interactions
are neglected. In simple terms it is sufficient to describe only one particle, 
embedded in an average potential generated by the other particles, 
in order to automatically describe the whole system.
In this case 
$S({\bf r_{1}},......{\bf r_{N}})=s({\bf r_{1}})+....s({\bf r_{M}})$.
 It must be noticed that, within the approximation done, 
  what we called the
{\bf one-particle Bohm potential},  
{\bf does not correspond to the Bohm potential 
for a system composed by only one particle}. In fact, if this was the case, 
Bohm equations could not be defined at the nodes of the electron 
wavefunction, instead
the definition of $\psi({\bf r})$ given in Eq.\ref{wf} with $\rho({\bf 
  r})$ defined by Eq.\ref{dr} suggests that it is unlikely to have a $\psi$
which contains zeros at least for dense systems.
The one-particle potential 
 is the simplest approximation that can be used for describing an
electronic system. Recently M.Kohout proposed another approach,
by redefining the wavefunction $\psi({\bf r},{\bf r'})$ (${\bf
  r'}$ indicates the remaining $N-1$ particles of an $N$ -particle
system when one focuses the attention on particle ${\bf r}$):
\begin{equation}
\psi({\bf r},{\bf r'})=\chi({\bf r},{\bf r'})e^{i\hbar S({\bf r},{\bf r'})}=
\phi({\bf r})\cdot \beta({\bf r'|r})e^{i\hbar S({\bf r},{\bf r'})}.
\label{cf}
\end{equation}
$\phi({\bf r})$ is the marginal function while $\beta({\bf r'|r})$ is
the conditional function, conditionally dependent on the coordinate
${\bf r}$.
These two functions satisfy the following normalization conditions:
\begin{equation}
\int_{\Omega}|\phi({\bf r})|^{2}d{\bf r}=1
\label{prob1}
\end{equation}
and
\begin{equation}
\int_{\Omega}|\beta({\bf r'|r})|^{2}d{\bf r'}=1; \forall {\bf r}.
\label{prob2}
\end{equation}
By using this factorization, he obtains a formal expression of the 
potential $Q({\bf r},{\bf r'})$ as a sum of the one-particle 
Bohm potential and a conditional multi-particle potential
$Q_{cond}({\bf r}|{\bf r'})=-\frac{\hbar}{2m\beta({\bf
    r'|r})}\sum_{i=1,N}\nabla_{i}^{2}\beta({\bf r'|r})$.
This expression bares a rather difficult problem: {\bf to find a reasonable 
expression for} $\beta({\bf r'|r})$. In this work we will circumvent
this problem by using a suitable expression for $\chi({\bf r},{\bf r'})$
which recovers the properties of the factorized wavefunction introduced by
Kohout, and leads to an iterative procedure for $Q({\bf r},{\bf r'})$;
this is reported in the next sections.
\section{Many-particle density and Bohm's wavefunction.}  
As anticipated in previous sections our final aim is to develop a 
procedure for obtaining a Bohm's potential where many-particle effects
are somehow incorporated. For this purpose, we redefine the
wavefunction of the system by extending the form $\psi({\bf
  r})=\phi({\bf r})e^{i\hbar s({\bf r})}$ for one-particle to $\psi({\bf
  r_{1},r_{2}...r_{M}})=\phi({\bf  r_{1},r_{2}...r_{M}})e^{i\hbar S({\bf
    r_{1},r_{2}...r_{M}})}$, an $M$-particle wavefunction in an
$N$-particle system, with $M\leq N$. We should also require $\phi$ to
be antisymmetric and $S$ to be symmetric 
with respect to the $M!$ possible pair permutations of
the $M$ particles; this requirement will preserve the antisymmetry of $\psi$. 
Later we will show
that the symmetry of $S$ corresponds to the fundamental 
physical property of indistinguishable particles. 
In the next sections, to remind
the analogy with the one-particle case we will identify ${\bf r}$
with ${\bf r}_{1}$, thus $\phi({\bf  r_{1},r_{2}...r_{M}})=\phi({\bf
  r,r_{2}...r_{M}})$.
This basically means that given $N$ particles, our system is
characterized by (or alternatively, we are interested in considering) 
$M$-particle interactions which produce observable effects on the average behavior 
of the system, thus $M$ particles must be treated explicitly; we can describe those effects by considering the $M$-th 
approximation, where $M=1,...N$, the case $M=1$ is the trivial 
one-particle case, $M=2$ counts two-particle effects etc.etc..
For the moment we are not interested in $S({\bf
    r_{1},r_{2}...r_{M}})$ whose role will be clear later on. Let us
  focus on $\phi({\bf  r,r_{2}...r_{M}})$.
In the one-particle case 
$N\cdot|\phi({\bf  r})|^{2}= \rho({\bf r})=N\int_{\Omega^{(N-1)}}|\psi({\bf r,
  r_{2},r_{3}......r_{N}})|^{2} 
d{\bf{r_{2}}}d{\bf{r_{3}}}....d{\bf{r_{N}}}=\rho({\bf
  r})$,
i.e. the average electron density of $N$ indistinguishable
particles projected on the real ($3$-dimensional) space.
In analogy we can define an M-particle electron density as :
\begin{equation}
N|\phi({\bf r....r_{M}})|^{2}=\rho({\bf r....r_{M}})=  N\int_{\Omega^{(N-M)}}|\psi({\bf r,
  r_{2}..r_{M},r_{M+1}......r_{N}})|^{2} d{\bf{r_{M+1}}}....d{\bf{r_{N}}}
\end{equation}
where, in analogy to the one-particle case, one has:
\begin{equation}
\int_{\Omega^{M}}|\phi({\bf
  r....r_{M}})|^{2}d{\bf{r}}....d{\bf{r_{M}}}=1.
\end{equation}
This form of $\phi({\bf r,r_{2}...r_{M}})$ 
satisfies the requirements of the factorization
in marginal and conditional part as defined by Kohout:
\begin{equation}
\psi({\bf r,r_{2},....r_{M}})=\phi({\bf r})\cdot\frac
{\phi({\bf r,r_{2}...r_{M}})}{\phi({\bf r})}e^{i\hbar S({\bf
    r,r_{1}...r_{M})}}.
\label{formally}
\end{equation}
In fact the conditional function of Eq.\ref{cf},
$\beta({\bf r'|r})$, can be written as: $\beta({\bf r'|r})=\frac{\phi({\bf
    r,r_{2}....r_{M}})}{\phi({\bf r})}$ and (as can be easily verified)
one obtains $\int_{\Omega^{M}}|\beta({\bf r'|r})|^{2}d^{M}{\bf r'}=1; 
\forall {\bf r} 
\in \Omega$, where ${\bf r}'={\bf r_{2},r_{3},...r_{M}}$.
This kind of factorization was already considered by Hunter
\cite{hunter} who, once more, underlines that 
the nature of $\phi({\bf r})$, as a marginal probability density
function, makes unlikely the existence of zeros. 
In terms of probability $\frac{\phi({\bf
    r,r_{2}....r_{M}})}{\phi({\bf r})}$ is interpreted 
as the square root of the classical expression for the conditional
probability density. 
The conditional probability density, i.e. the probability density 
 for $M-1$ particles given the position of the particle
${\bf r}$, is written as the
probability density of the particles ${\bf r,r_{2}.....r_{M}}$, ($|\phi({\bf
    r,r_{2}....r_{M}})|^{2}$), divided
by the probability density of particle ${\bf r}$, ($|\phi({\bf r})|^{2}$).
In general the mathematical structure of quantum 
mechanics leads to a non-commutative probability theory which coincides
with the classical one only in case we treat commuting
spaces \cite{cassinelli}. Here for a {\it commuting
space} we intend a set of variables representing physical quantities,
such as positions, which do commute.
In the language of quantum mechanics this 
means for example that position operators of the $M$-particles 
commute, i.e. a measurement 
of the position of particle $1$ does not influence
the measurement of the position of particle $2$. The same example does
not hold in the space of spins, however 
since we have restricted our analysis to 
spinless systems, we can apply the rules of classical probability.
In Ref.\cite{cassinelli} is reported a study about
the possibility of
defining a quantum correction which takes into account a
non-commutative probability  for the conditional function. 
Once we have formally defined the many-electron wavefunction via
Eq.\ref{formally}, we 
can determine a reasonable initial guess for Bohm's potential, by
defining a reasonable $\phi({\bf r,r_{2}....r_{M}})$ : 
\begin{equation}
Q({\bf r,r_{2}....r_{M}})=-\frac{\hbar}{2m\phi}\sum_{i=1,M}\nabla^{2}_{i=1,M}\phi.
\end{equation} 

\section{Bohm's dynamical system for $M$ non-independent particles}
In the previous section we defined, in analogy to the one-particle
case, an M-particle wavefunction for an $N$-particle system.
Clearly the larger the value of $M$ the more difficult the
determination of the wavefunction, although in many cases 
two or three-particle effects may be enough for a basic
understanding of some physical properties. So far by defining
the many-particle wavefunction we simply gave a first approximation
for the potential $Q$, within, for example, Slater orbitals or similar 
appropriate approximations (e.g. plane-waves expansion ). {\bf The idea
for an effective
coupling of $M$ electrons is to insert an approximation of $Q$, as
an initial guess, in a
system of equations which do couple the $M$ electrons}. This set of equations
is obtained by inserting the M-particle wavefunction in Schr\"{o}dinger
equation and repeating the procedure followed for the one-particle
case within Bohm's framework as it is shown next. 
In fact, by inserting $\psi({\bf r, r_{2},...r_{M}})$ in  Schr\"{o}dinger
equation for the stationary case, similarly to Eq.\ref{genmot2}, one obtains:
\begin{equation}
\sum_{i={\bf r,r_{M}}}\frac{[\nabla_{i}S({\bf r,...r_{M}})]^{2}}{2m}=-V({\bf
  r,...r_{M}})-Q({\bf r,...r_{M}})
\label{multieq}
\end{equation}
where $-V({\bf r,...r_{M}})$ is the electrostatic potential
experienced by each single particle.
In this form, Eq.\ref{multieq} is not very useful thus we should find
a way for a simplification which can be based on physical well founded
hypothesis. The first thing to notice is that $\nabla_{r_{i}}S({\bf
  r,...r_{M}})={\bf v}_{r_{i}}({\bf r,...r_{M}})$ represents, within
Bohm's framework, the velocity field of particle ${\bf r}_{i}$ which
depends upon (or alternatively, is influenced by) the positions of the 
other $M-1$ particles as well. 
We can also notice that  $-V({\bf
  r,...r_{M}})$ can be represented as the electrostatic potential per
particle and can be expressed in good approximation as:
\begin{equation}
V({\bf
  r,...r_{M}})=\sum_{i=1,M}\frac{1}{N}\left[\frac{1}{2}\int_{\Omega}\frac{\rho({\bf w})}{{\bf r}_{i}-{\bf w}}d{\bf w}\right].
\label{electro}
\end{equation}
In Eq.\ref{electro}, $\rho({\bf w})$ is the one-particle electron
density, obtained by integrating the $M$-particle density over $M-1$
variables. The meaning of Eq.\ref{electro} is that in a system of
indistinguishable particles the average electrostatic potential experienced by 
one-particle is the same as that experienced by another particle. The ''average'' character is obtained by 
considering the one-particle electron density instead of the
$M$-particle one.
 The use of a simpler 
rather than a a more complicated expression for $V$ can be justified by the
fact that we know how to
express electrostatic properties and we know that Eq.\ref{electro} is
reasonable. What is unknown are the  
quantum effects represented by $Q$, for this reason we can take known 
quantities in their simplest approximation as far as they are known to be 
reasonable. In any case the important fact is that
following  Eq.\ref{electro} one can write:
\begin{equation}
V({\bf r,..r_{M}})=V({\bf r})+V({\bf r_{2}})+....+V({\bf r_{M}}).
\label{split}
\end{equation}
The expression for $V({\bf r,..r_{M}})$ given in  Eq.\ref{split} is
particularly useful for simplifying Eq.\ref{multieq}.
In fact Eq.\ref{multieq} can be written as :
\begin{equation}
\sum_{i={\bf r,r_{M}}}\frac{[\nabla_{i}S({\bf
  r,...r_{M}})]^{2}}{2m}=-\sum_{i=1,M}\left[V({\bf r}_{i})+\frac{1}{M}Q({\bf
    r,r_{2}...,r_{M}})\right]
\label{summul}
\end{equation}
where $\frac{1}{M}Q({\bf r,r_{2}...,r_{M}})$ is the quantum potential
per {\it interacting particle}. 
At this point we can proceed having in mind the following points:\\
Eq.\ref{summul} is a $3M$-dimensional 
 non linear partial differential equation and
we would like to have a solution which somehow recovers the dynamical
physics contained in Bohm's approach.
The natural simplification of Eq.\ref{summul} is to decompose it in
a system of $M$ one-particle equations coupled through $S({\bf
  r,r_{2}...r_{M}})$ and $Q({\bf r,r_{2}...,r_{M}})$, where each equation
describes the squared modulus of the velocity field of a particular electron $\nabla_{i}S({\bf
  r,...r_{M}})$ subject to a potential {\it per particle} $\left[V({\bf r}_{i})+\frac{1}{M}Q({\bf
    r,r_{2}...,r_{M}})\right]$. Clearly the
sum of solutions of single equations is also a solution for the
initial one. In mathematical
terms this kind of
solution represents only one possible way (a particular solution)
, while in physical terms
this represents  the direct extension of Bohm's one-particle dynamical
description to a system of $M$-particle non-independent
electrons . In this way we obtain a set of equations coupling 
the $M$ interacting electrons as we postulated at the beginning of this 
section.The 
$M$-equation system takes the form:
\begin{equation}
\begin{array}{ll}
{(\bf 1)}[{\bf v}_{r}({\bf r,...r_{M}})]^{2}=\frac{[\nabla_{r}S({\bf
r,...r_{M}})]^{2}}{2m}=-\left[V({\bf r})+\frac{1}{M}Q({\bf
    r,r_{2}...,r_{M}})\right]\\
{(\bf 2)}[{\bf v}_{r_{2}}({\bf r,...r_{M}})]^{2}=\frac{[\nabla_{r_{2}}S({\bf
r,...r_{M}})]^{2}}{2m}=-\left[V({\bf r_{2}})+\frac{1}{M}Q({\bf
    r,r_{2}...,r_{M}})\right]\\
{(\bf 3)}............................................................\\
{(\bf 4)}............................................................\\
{(\bf 5)}............................................................\\
{(\bf 6)}............................................................\\ 
    .\\                                      
    .\\
    .\\
{(\bf M)}[{\bf v}_{r_{M}}({\bf r,...r_{M}})]^{2}=\frac{[\nabla_{r_{M}}S({\bf
r,...r_{M}})]^{2}}{2m}=-\left[V({\bf r_{M}})+\frac{1}{M}Q({\bf
    r,r_{2}...,r_{M}})\right].
\end{array}
\label{sys1}
\end{equation}
In the next section we show that, by using Eq.\ref{sys1},  
a self-consistent procedure for $Q({\bf
    r,...r_{M}})$ can be determined.
\section{Self-Consistent iterative procedure for $Q({\bf
    r,...r_{M}})$}
\begin{figure}
\centerline{\psfig{figure=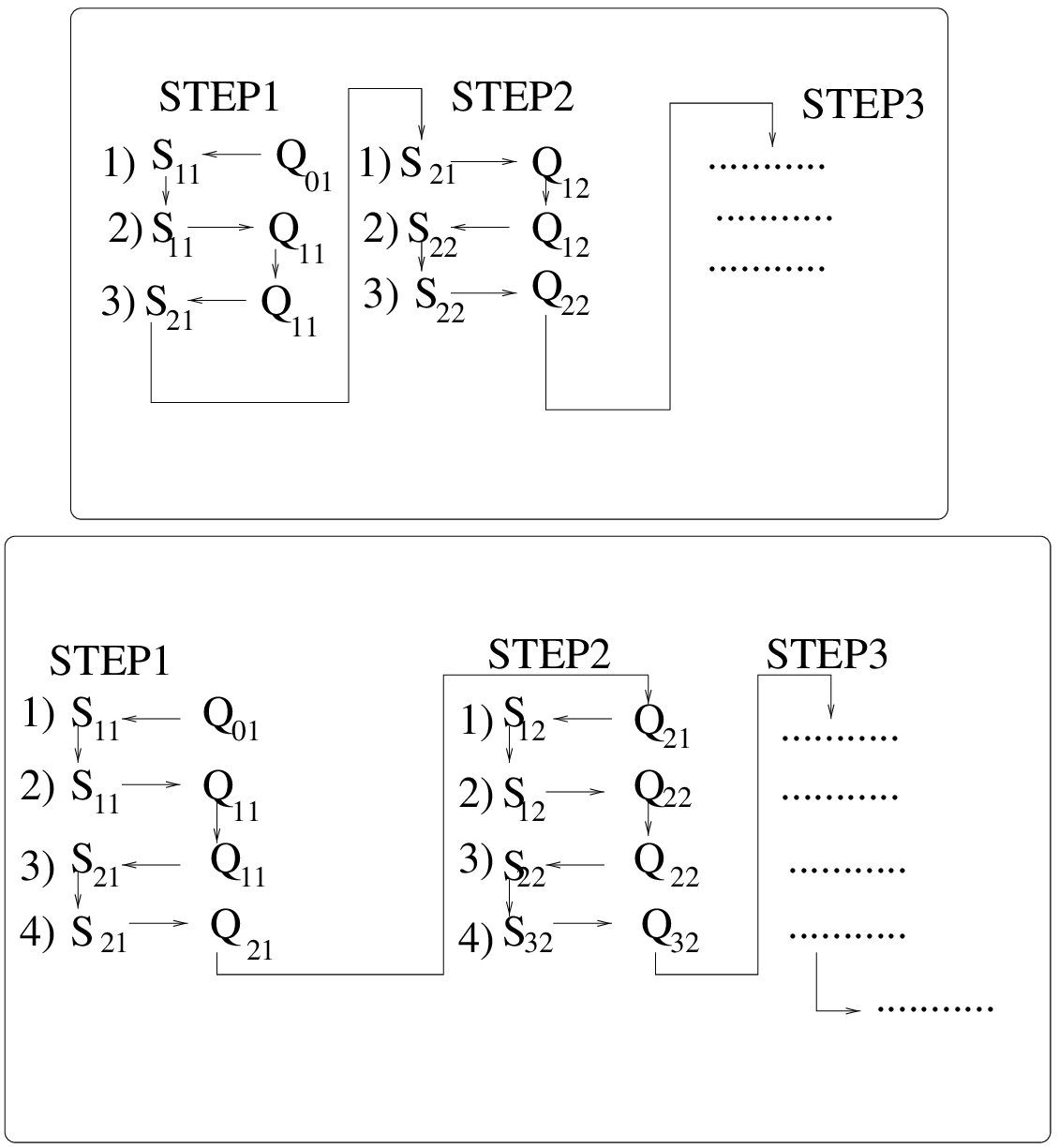}}
\caption{The figure represents the pictorial
    illustration
of the self-consistent iterative procedure shown in this work. 
For STEP we intent the process of going through all the 3 (4) equations.
Next by using the final $Q$ or $S$, obtained at the first STEP, 
it is possible to start a second STEP. $S_{ij}$ ($Q_{ij}$) is the
function $S$ calculated at different stages of the procedure. $i$
corresponds to the number of equations solved within the system
, $i=0$ corresponds to the starting (initial guess) expression for
$Q$, $j$ indicates the number of global iterations (STEP).
\label{fig}}
\end{figure}

The basic idea is to start from the approximative form of $Q=Q_{0}({\bf
    r,...r_{M}})$ obtained via the $M$-particle density within a
  Slater orbitals or similar approach and then introduce it in equation ${\bf
    (1)}$ of system \ref{sys1}. Next by solving the first order non
  linear partial differential equation for $S({\bf r,...r_{M}})$ with
  respect to ${\bf r}$ one find an ''initial''  value for $S({\bf
    r,...r_{M}})=S_{1}({\bf r,...r_{M}})$. At this point $S_{1}({\bf
    r,...r_{M}})$ is substituted in equation ${\bf (2)}$ and leads
  to a new value of $Q$, $Q_{1}({\bf
    r,...r_{M}})= \frac{[\nabla_{r_{2}}S_{1}({\bf
r,...r_{M}})]^{2}}{2m}+\left[V({\bf r_{2}})\right]$. The next step consists in
substituting $Q_{1}$ in equation ${\bf (3)}$ and solve it (as for equation
${\bf (1)}$ ) with respect to
${\bf r}_{3}$ to find a new $S({\bf r,...r_{M}})=S_{2}({\bf
  r,...r_{M}})$, the procedure is repeated until 
the convergence of $Q$ meets a given criterion of acceptance. It must be noticed that the procedure is
iterative in the sense that one should go through all the $M$-equations
of the system and then take the final $Q$ or $S$ for repeating the
process by going through the $M$ equations again as is illustrated 
for the case $M=3$ and $M=4$ in Fig.\ref{fig}.
Finally $Q({\bf r,r_{1}......r_{M}})$ can be integrated with respect
to $M-1$ variable and so be reduced to a one particle potential where
the many-particles effects are integrated out (averaged).
However, in practical terms this procedure is not straightforward since
a first order non linear 
partial differential equations must be solved, for example 
with respect to ${\bf r}=(x,y,z)$ at the very first step:
\begin{equation}
\frac{1}{2m}\left[\left(\frac{\partial S({\bf r,...r_{M}})}{\partial x}\right)^{2}+
\left(\frac{\partial S({\bf r,...r_{M}})}{\partial y}\right)^{2}+
\left(\frac{\partial S({\bf r,...r_{M}})}{\partial z}\right)^{2}\right]=-
\left[V({\bf r})+\frac{1}{M}Q_{0}({\bf
    r,...r_{M}})\right]
\label{partdiff}
\end{equation}
and formally a solution of the following form should be obtained:
\begin{equation}
S({\bf r,...r_{M}})=f({\bf r,r_{2},...r_{M}})+G({\bf
  r_{2},...r_{M}})+const.
\label{sss}
\end{equation}
$G({\bf r_{2},...r_{M}})$ is a function constant with respect to ${\bf 
  r}$ and $const$ is a constant which can be neglected 
since we are interested in 
gradients of $S$. 
$G({\bf r_{2},...r_{M}})$  
cannot be obtained by solving a single equation
since in principle one should solve the whole system of equations and
find the global solution (basically $G({\bf r_{2},...r_{M}})$ would be an
artifact of the procedure). On the other hand to find a global
solution would represent a formidable, if not 
impossible, task. We need a well founded  approximation which allow us
to remove the artifact represented by the function $G({\bf
  r_{2},...r_{M}})$.
At this point we notice that there exists a fundamental physical property 
of the system which can reasonably solve this problem.
In fact $S({\bf r},{\bf r}_{1}.....{\bf r}_{M})$ {\it must} be
symmetric under any permutation of the $N$ particles. This  
fundamental property is a direct consequence
of the fact that the particles are indistinguishable and implies that the
velocity field of one particle can be obtained from that of another via
particle permutation (see Appendix).

The symmetry of the total wavefunction can be preserved by taking
$\phi({\bf r,r_{1},....r_{M}})$ antisymmetric with respect to any
pair exchange of the $M$ particles.
The property of symmetry for $S$ 
allows us to make the iterative procedure possible and
physically reasonable. In fact we solve Eq.\ref{partdiff} with respect 
to ${\bf r}$ and obtain $S({\bf r})$ as:
\begin{equation}
S_{1}({\bf r,...r_{M}})=f({\bf r,r_{2},...r_{M}})
\end{equation}
formally neglecting, for the moment, the part $G({\bf
  r_{2},...r_{M}})$ of Eq.\ref{sss}.
Then in order to obtain a global solution which satisfies the
permutation (symmetry) criterion we write the solution for $S$
at the first iterative step as :
\begin{equation}
S_{1tot}({\bf
  r,r_{2},...r_{M}})=\frac{1}{M!}\sum_{i}P_{i}S_{1}({\bf r,...r_{M}})
\end{equation}
where $\sum_{i}P_{i}$ is the sum over all the possible $M$-coordinate
pair permutations and $P_{i}$ is the permutation operator for the $i$-th 
permutation. This procedure, by making a fully symmetric solution $S$,
does not require to consider the term $G({\bf r_{2},...r_{M}})$ which
would formally come from a direct mathematical approach.
This point can be considered from an alternative point of view
; we extract a solution of a
multi-variable function problem by considering only one equation
instead of considering all the equations.
Such a solution is valid for the variable ${\bf r}$
(integration variable) as long as the other variables are considered 
as constants (fixed at a particular parametric value).
Next, by using the physical condition of indistinguishable particles,
we can extend, reasonably well, this solution to the whole set of
variable by making the global solution a fully symmetric function of
the whole set of coordinates. We can then go to the next equation 
to find $Q$, make it fully symmetric with the same procedure we used
before for $S$ and go to the next equation, again solve for $S$, make it
symmetric and proceed on in the same way. 
A formal argument for the symmetry of $S$ is given in Appendix.    
Technically the major problem is the fact
that in general to solve a non
linear partial differential equation in three dimensions 
is not an easy task, however 
 Eq.\ref{partdiff} is the well known equation 
of the eikonal in an anisotropic media. In this case the boundary
condition for $S$
must be assigned  when the Schr\"{o}dinger equation  is formulated
and should depend on the particular nature of the system under consideration.  
The problem of the solution of this equation goes beyond the purposes
of this work, however  there exists a massive
list of references dealing with this problem  
for both mathematical properties and numerical techniques (see for 
example \cite{eik1,eik2,eik3,eik4,eik5,eik6,eik7,eik8,eik9} and references
therein). In particular the fast sweeping algorithms of Ref.\cite{eik6},
and the robust algorithms for multidimensional Hamilton-Jacobi
equations of Refs.\cite{eik7,eik8,eik9} 
represent an extremely useful approach to a
computer implementation of this procedure. In particular these latter algorithms,
although dealing with the more general non stationary problem, in
principle can be adapted to the stationary case and in general would
make it possible also to extend the procedure to non stationary cases.

\section{Conclusions}
We propose a general method to treat Bohm's potential in its $3N$-dimensional 
rigorous form. Inevitably, there are several physical approximations
that one should accept; we use a spinless system, thus we can apply
the concept of classical probability to a quantum system in a
stationary state. This choice offers technical advantages but is not
physically obvious; in case the spin variables are explicitly
considered, the basis of our procedure remains valid, but we should
find an opportune way to define  a more general wavefunction where the
conditional function is determined by non-commutative probability
principles as suggested in Ref.\cite{cassinelli}. 
The separation of the many-electron equation in
single but mutually dependent one-particle equations is not unique
from a rigorous mathematical point of view but it is based on a
reasonable physical approximation.
From a technical-mathematical point of view the major problem is
the solution of the non-linear partial differential equations for $S$
(eikonal equation).
As it is commented in several textbooks of quantum mechanics (see for
example \cite{sakurai}), to solve 
this equation is not an easy
task, however the literature relative to the solution of the
eikonal equation is sufficiently large and several methods are available.
For small $M$ (e.g. $2,3$) it should be possible to apply the procedure with a
reasonable effort; moreover there exists systems which can be
described with simplified unidimensional models such as
electrons in a one dimensional wire or an isolated atom
considered spherically symmetric.
Other simple examples are those where it is possible to proceed to a
separation of variables
(i.e.$f(x_{i},x_{j},y_{i},y_{j},z_{i},z_{j})=X(x_{i},x_{j})+Y(y_{i},y_{j})+Z(z_{i},z_{j})$
or
$f(x_{i},x_{j},y_{i},y_{j},z_{i},z_{j})=X(x_{i},x_{j})Y(y_{i},y_{j})Z(z_{i},z_{j})$) 
in this case the eikonal equation is reduced to independent ordinary
first order differential equations which can be easily solved.
The simple cases listed above can be used as a 
 first approximation of more complicated systems.
As stated before, it is important to underline that the 
intention of this work is that of proposing a procedure to properly
treat and use Bohm potential in fields where it is currently (and
extensively) employed for practical applications.
We do not claim that this procedure is computationally or
 methodologically more
convenient than others in solving  {\bf general} 
many-electron problems , however, this method 
may {\bf also} represent a  complementary {\bf theoretical} 
approach to the standard ones used in
current research; due to the deterministic interpretation
of Quantum Mechanics, on which the method is based, effects 
which cannot be describe by standard Hartree-Fock or DFT may be
revealed. This issue, anyway, involves a much deeper analysis 
which goes beyond the purpose of this work and will be possibly
treated elsewhere.

\section{Acknowledgments}
I would like to thank Dr.M.Kohout, Prof.T.Vilgis and Dr.E.Cappelluti 
for a critical reading of the manuscript and helpful suggestions.
\section{Appendix}
For simplicity let us consider a two-particle system (${\bf r,r'}$).
Let us suppose that from Eq.\ref{partdiff} for (${\bf r,r'}$), 
integrated with respect to ${\bf r}$, the following solution is found:
\begin{equation}
f({\bf r,r'})=g({\bf r,r'})+h({\bf r})+l({\bf r'})
\label{rrp}
\end{equation}
where $g({\bf
  r,r'})=g({\bf r',r})$ and $l({\bf r'})$ is unknown since it
represents the 
constant (with respect to ${\bf r}$) obtained by solving the equation
with respect to  ${\bf r}$.
From Eq.\ref{rrp} it follows that the formal expression for 
the velocity field for the
particle  ${\bf r}$ is 
\begin{equation}
{\bf v}_{\bf r}({\bf r,r'})=\nabla_{\bf r}[g({\bf r,r'})+h({\bf r})]
\end{equation} 
this means
\begin{equation}
{\bf v}_{\bf r}({\bf r,r'})={\bf v}_{1}({\bf r,r'})+
{\bf v}_{2}({\bf r})
\end{equation}
while
for particle ${\bf r'}$ we have:
\begin{equation}
{\bf v}_{\bf r'}({\bf r,r'})=\nabla_{\bf r'}[g({\bf r,r'})+l({\bf r})]={\bf v}_{1}({\bf r',r})+{\bf v}_{3}({\bf
  r'}).
\end{equation}
Because the particles are indistinguishable 
one must obtain ${\bf v}_{\bf r'}({\bf r',r})$ 
from ${\bf v}_{\bf r}({\bf r,r'})$ by permutation of (${\bf r,r'}$) thus 
the unknown function $l({\bf r'})$ must coincide with $h({\bf r'})$.
As one can easily see, by taking $f({\bf r,r'})=g({\bf r,r'})+h({\bf
  r})$
and making it symmetric with respect to the permutation $({\bf r,r'})\to ({\bf
  r',r})$, we reach the same result for $l({\bf r'})$.
In the most general case $g({\bf r,r'})$ is not necessarily equal to 
$g({\bf r',r})$, however the same procedure (and principles) can be
applied with the only difference that $g$ must be also made symmetric,
i.e. $g_{fin}({\bf r,r'})= \frac{g({\bf r,r'})+g({\bf r',r})}{2}$ and 
$h_{fin}({\bf r,r'})=\frac{h({\bf r})+h({\bf r'})}{2}$. 
In case of more than 2 particles the same argument can be used by
taking into account all possible particle permutations.

\newpage

\end{document}